\documentclass[a4paper]{spie}  
\usepackage{amsmath}
\usepackage{xfrac}
\usepackage[]{graphicx}

\title{Limitation in thin-film detection with\\transmission-mode terahertz time-domain spectroscopy}

\author{Withawat Withayachumnankul
\skiplinehalf
School of Electrical \& Electronic Engineering, The University of Adelaide,\\Adelaide, SA 5005, Australia \\
Faculty of Engineering, King Mongkut\rq{}s Institute of Technology Ladkrabang,\\Bangkok 10520, Thailand
}

\authorinfo{withawat@eleceng.adelaide.edu.au}
 
\begin{document} 
\maketitle 

\begin{abstract}
In transmission-mode terahertz time-domain spectroscopy (THz-TDS), the thickness of a sample is a critical factor that determines an amount of the interaction between terahertz waves and bulk material. If the interaction length is too small, a change in the transmitted signal is overwhelmed by fluctuations and noise in the system. In this case, the sample can no longer be detected. This article presents a criterion to determine the lower thickness boundary of a free-standing film that can still be detectable by free-space transmission-mode THz-TDS. The rigorous analysis yields a simple proportional relation between the sample optical length and the system SNR. The proposed criterion can help to decide whether an alternative terahertz thin-film sensing modality is necessary.
\end{abstract}


\section{Introduction}

Teraherz time-domain spectroscopy (THz-TDS) is an efficient tool for characterizing materials with broadband coherent terahertz radiation \cite{Wit07c}. A femtosecond optical laser is utilised in THz-TDS. Interaction of optical pulses with an emitter results in a burst of subpicosecond pulses spanning a range from a few hundred gigahertz to a few terahertz. The emitted pulses induce a local change in the electric field at the detector. A pump-probe detection scheme resolves the amplitude and phase of a terahertz pulse with a high SNR. In a transmission measurement, the generated terahertz pulses are changed in the amplitude and phase after interacting with a sample. These changes can be used to extract the complex optical constants, or related quantities, of the sample. 


For free-space transmission-mode THz-TDS, each material has an optimal thickness that yields the lowest measurement uncertainty in the optical constants \cite{Wit08c}. The noise floor clearly defines the maximum thickness of a sample that still results in the valid optical parameters \cite{Jep05}. For a thin sample, the measurement accuracy significantly reduces, because the signal variation due to fluctuations and noise in the system becomes relatively strong compared with the change in the amplitude and phase introduced by the sample. The limitation of thin-film \textit{characterization} exists where the film is so thin that the extracted parameters are no longer reliable. Likewise, \textit{detecting} the presence of a thin film reaches a limit when the sample is no longer distinguishable from the background material. Despite that, there is no explicit criterion for determining a lower boundary for the thickness of a sample under free-space THz-TDS measurement.

This article presents a mathematical analysis on the limitation of free-space THz-TDS for thin-film \textit{detection}. Initially, the thin-film condition is expressed in terms of optical constants and their standard deviations, which are caused by a variation in the signal amplitude. These standard deviations are then further expanded to yield the criterion of the detection limitation. It defines the minimal film thickness and refractive index as a function of the system SNR. The derivable condition is intuitive and useful as a rule of thumb in estimating the capability of a THz-TDS system. It also suggests when other advanced measurement techniques are necessary to detect particular samples. 


\section{Limitation for thin-film detection}

\begin{figure}[b]
\centering
\includegraphics{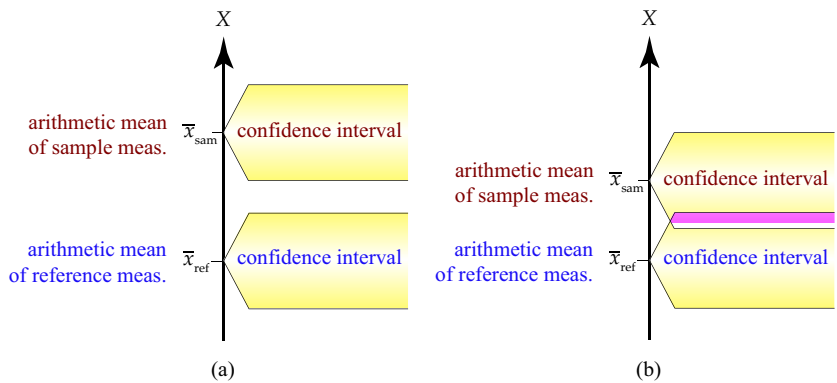}
\caption{Confidence intervals in the detection scheme. (a) The confidence intervals of the sample and reference measurements are separated. The presence of the sample is confirmed. (b) The two confidence intervals partially overlap. The presence of the sample is unclear.}
\label{fig:MK2_confidence_interval}
\end{figure}

In the context of the uncertainty analysis, the random error, which encompasses fluctuations and noise, is used to evaluate the confidence interval in a measurement. This confidence interval \textit{localizes} or bounds the expected value, i.e., the value that is free from the random error. Therefore, in the sample detection scheme, if the confidence intervals of the sample and reference measurements are well separated, it can be inferred that the sample is unambiguously detected. Otherwise, if the two intervals get overlapped, it is possible that the two measurements share an identical expected value. Hence, in this case, the presence of the sample cannot be judged. Fig.~\ref{fig:MK2_confidence_interval} schematically illustrates both of the situations. It is worth nothing that the systematic error is irrelevant in the detection scheme, since it biases the sample and reference measurements equally. 

Mathematically, the confidence interval for any variable $x$ is defined as \cite{GUM95}
\begin{eqnarray}\label{eq:MK_confidence_interval}
\bar{x} - \frac{s_x}{\sqrt{N}} \leq \mu_x \leq \bar{x} + \frac{s_x}{\sqrt{N}} \;,
\end{eqnarray}
where $\bar{x}=\sum_{l=1}^N x_l/N$ is the arithmetic mean from $N$ measurements, $s_x$ is the empirical
standard deviation, and $\mu_x$ is the expected value. From Eq.~\ref{eq:MK_confidence_interval} and the discussion in the previous paragraph, it can be established that the sample with optical constants $n_s(\omega)+j\kappa_s(\omega)$ can be distinguished from the reference material with $n_0+j\kappa_0$ by terahertz waves if any of these two criteria is satisfied:
\begin{subequations}\label{eq:MK_criteria}
\begin{eqnarray}
\Delta n(\omega) &>&\sqrt{
\frac{s_{n,E_{\rm sam}}^2}{N_{E_{\rm sam}}}
+\frac{s_{n,E_{\rm ref}}^2}{N_{E_{\rm ref}}}
+\frac{s_{n,l}^2}{N_l}} \;, \quad \text{or}\\
\Delta\kappa(\omega) &>& \sqrt{
\frac{s_{\kappa,E_{\rm sam}}^2}{N_{E_{\rm sam}}}
+\frac{s_{\kappa,E_{\rm ref}}^2}{N_{E_{\rm ref}}}
+\frac{s_{\kappa,l}^2}{N_l}}
 \;,
\end{eqnarray}
\end{subequations}
where $\Delta n(\omega) = \bar{n}_s(\omega)-n_0$ and $\Delta\kappa(\omega)=\bar{\kappa}_s(\omega) -\kappa_0$; $s_{\{n,\kappa\},E_{\rm sam}}^2$, $s_{\{n,\kappa\},E_{\rm ref}}^2$, and $s_{\{n,\kappa\},l}^2$ are the empirical variances of the sample optical constants caused by the sample and reference amplitude variations, and thickness variation, respectively; $N_{E_{\rm sam}}$ and $N_{E_{\rm ref}}$ are the numbers of measurements for the sample and reference signals, respectively; and $N_l$ is the number of sample thickness measurements. Eq.~\ref{eq:MK_criteria} means that the difference between the sample refractive index or extinction coefficient and that of the reference material must be larger than its combined standard deviation, which is mainly contributed by two sources of random error, i.e., signal noise and thickness measurement \cite{Wit08}. 

Only the amplitude-related variances, $s_{\{n,\kappa\},E_{\rm sam}}^2$ and $s_{\{n,\kappa\},E_{\rm ref}}^2$ can be considered, provided that the error in thickness measurement is relatively easy to deal with or becomes irrelevant if the detection is based on the time-domain or frequency-domain data. Furthermore, $s_{\{n,\kappa\},E_{\rm sam}}^2$ and $s_{\{n,\kappa\},E_{\rm ref}}^2$ can be combined if the numbers of sample and reference measurements are equal. Hence, from the given assumptions, Eq.~\ref{eq:MK_criteria} can be simplified to
\begin{subequations}\label{eq:MK_criteriax}
\begin{eqnarray}\label{eq:MK_criteriax1}
\Delta n(\omega) &>& \frac{s_{n}(\omega)}{\sqrt{N}} \;, \\\label{eq:MK_criteriax2}
\Delta\kappa(\omega) &>& \frac{s_\kappa(\omega)}{\sqrt{N}} \;,
\end{eqnarray}
\end{subequations}
where $s^2_{\{n,\kappa\}}(\omega)= s_{\{n,\kappa\},E_{\rm sam}}^2+ s_{\{n,\kappa\},E_{\rm ref}}^2$ and $N=N_{\rm sam}=N_{\rm ref}$. Typically, it can be shown that
\begin{subequations}\label{eq:MK_nlk}
\begin{eqnarray}
\Delta n(\omega)  &>&  \Delta\kappa(\omega) \;,\quad\text{and}\\
s_n(\omega) &\le& s_\kappa(\omega)\;.
\end{eqnarray}
\end{subequations}
The conditions in Eq.~\ref{eq:MK_nlk} implies that the criterion in Eq~\ref{eq:MK_criteriax1} is less strict than Eq~\ref{eq:MK_criteriax2}. As discussed earlier, to detect the presence of the sample, satisfying only one criterion, here Eq.~\ref{eq:MK_criteriax1}, is enough.


From the relation between the standard deviation of the refractive index and that of the reference phase in Eq.~\ref{eq:ax_stddev1}, the criterion for sample detection given in Eq.~\ref{eq:MK_criteriax1} can be expressed as
\begin{subequations}\label{eq:MK_criteria2}
\begin{eqnarray}
\frac{\omega l}{c}\Delta n(\omega)  &>&  \sqrt{\frac{2}{N}}  s_{\arg(E_{\rm ref})}(\omega)\;.
\end{eqnarray}
\end{subequations}
Here, $l$ is the sample thickness, and $s_{\arg(E_{\rm ref})}(\omega)$ is the standard deviation in the phase of the reference spectrum. This equation is the ultimate criterion for any free-standing sample to be detectable by normal transmission-mode THz-TDS. It means that a phase change introduced to the terahertz pulse by a sample must be larger than the scaled standard deviation of the reference phase. This standard deviation $s_{\arg(E_{\rm ref})}(\omega)$ can be obtained from a set of repeated time-domain measurements by using Eq.~\ref{eq:variance_of_phase3} or the Monte Carlo method. Increasing the number of measurements $N$ does not reduce the minimum sample thickness, but on the contrary accentuating $s_{\arg(E_{\rm ref})}(\omega)$ due to the long-term laser instability \cite{Wit08}.

\section{Conclusion}

In this article, a limitation in thin-film detection using ordinary transmission-mode THz-TDS is derived on the basis of the uncertainty analysis. Provided that the standard deviation of the terahertz signal is known in priori, the proposed criterion can be used to estimate the minimal thickness and refractive index of a free-standing film that can be detected with a free-space THz-TDS measurement. In future work, a limitation in thin-film characterization and a limitation in the case of substrate-based films will be analyzed. The proposed criterion will be validated through experiments with different types of films.


\appendix

\section{Standard deviation in optical constants}\label{sec:derivation_stddev}

This section shows a derivation of the relation between the variance in the optical constants and the variance in the terahertz magnitude and phase. From the amplitude in the time domain, the variance propagates to the frequency domain via Fourier transform. The amplitude and phase variances of the reference and sample spectra then produce the variance of the transfer function. Eventually, the variance appears at the optical constants. The following derivation is adapted from the existing detailed analysis \cite{Wit08}

The discrete Fourier transform of a time-resolved signal, $E(k)$, is given as
\begin{eqnarray}
	E(\omega)&=&\sum_{k}E(k)\exp(-j\omega k\tau)\;,
\end{eqnarray}
where $k$ is the temporal index, $\tau$ is the sampling interval, and $E(\omega)=E_{\rm r}(\omega)+jE_{\rm i}(\omega)$. Assuming that the amplitude at each time sample is statistically independent from the
amplitude at other time samples, the variances of the real and imaginary parts $s^2_{E_r}(\omega)$ and $s^2_{E_i}(\omega)$, and their covariance $s_{E_rE_i}(\omega)$ are, respectively \cite{Fornies97},
\begin{subequations}\label{eq:variance_re_im}
\begin{eqnarray}
	s_{E_{\rm r}}^2(\omega)&=&\sum_{k}\cos^2(\omega k\tau)s_{E}^2(k)\;,\\
	s_{E_{\rm i}}^2(\omega)&=&\sum_{k}\sin^2(\omega k\tau)s_{E}^2(k)\;,\\
s_{E_{\rm r}E_{\rm i}}(\omega)&=&-\sum_{k}\sin(\omega k\tau)\cos(\omega k\tau)s_{E}^2(k)\;,
\end{eqnarray}
\end{subequations}
where $s_{E}^2(k)$ is the variance of the time-domain signal $E(k)$. The magnitude and phase of the terahertz signal, determined from the real and imaginary parts of the complex spectrum, are, respectively,
\begin{subequations}
\begin{eqnarray}
	\ln|E(\omega)|&=&\frac{1}{2}\ln\left(E_{\rm r}^2(\omega)+E_{\rm i}^2(\omega)\right)\;,\\
	\arg( E(\omega))&=&\arctan(E_{\rm i}(\omega)/E_{\rm r}(\omega))\;.
\end{eqnarray}
\end{subequations}
The variances in the magnitude and phase are therefore given by, respectively,
\begin{subequations}
\begin{eqnarray}\label{eq:variance_of_magnitude3}
	s_{\ln|E|}^2(\omega)&=&\frac{1}{|E(\omega)|^4}\sum_{k}\Re^2[E(\omega)\exp(j\omega k\tau)]s^2_{E}(k)\;,\\\label{eq:variance_of_phase3}
	s_{\arg( E)}^2(\omega)&=&\frac{1}{|E(\omega)|^4}\sum_{k}\Im^2[E(\omega)\exp(j\omega k\tau)]s^2_{E}(k)\;,
\end{eqnarray}
\end{subequations}
where $\Re^2$ and $\Im^2$ denote the square of the real and of imaginary parts, respectively.
The transfer function of a sample is calculated from the reference and sample complex spectra as follows
\begin{subequations}
\begin{eqnarray}\label{eq:transfer_magnitude}
	\ln|H(\omega)|&=&\ln|E_{\rm sam}(\omega)| - \ln|E_{\rm ref}(\omega)|\;,\\
	\label{eq:transfer_phase}
	\arg (H(\omega))&=&\arg( E_{\rm sam}(\omega))-\arg( E_{\rm ref}(\omega))\;.
\end{eqnarray}
\end{subequations}
The variances of Eq.~\ref{eq:transfer_magnitude} and \ref{eq:transfer_phase} are, respectively,
\begin{subequations}
\begin{eqnarray}\label{eq:variance_trans_magnitude}
	s_{\ln|H|}^2(\omega)&=&s_{\ln|E_{\rm sam}|}^2(\omega) + s_{\ln|E_{\rm ref}|}^2(\omega)\;,\\\label{eq:variance_trans_phase}
	s_{\arg (H)}^2(\omega)&=&s_{\arg( E_{\rm sam})}^2(\omega)+s_{\arg (E_{\rm ref})}^2(\omega)\;.
\end{eqnarray}
\end{subequations}
If the reference and sample measurements have a comparable amplitude, Eq.~\ref{eq:variance_trans_magnitude} and \ref{eq:variance_trans_phase} can be simplified to
\begin{subequations}\label{eq:variance_trans_simplified}
\begin{eqnarray}
	s_{\ln|H|}^2(\omega)&=&2s_{\ln|E_{\rm ref}|}^2(\omega)\;,\\
	s_{\arg (H)}^2(\omega)&=&2s_{\arg (E_{\rm ref})}^2(\omega)\;.
\end{eqnarray}
\end{subequations}
The refractive index and the extinction coefficient are evaluated from the transfer function via
\begin{subequations}
\begin{eqnarray}\label{eq:index2}
	n_s(\omega)&=&n_0-\frac{c}{\omega l}\angle{H(\omega)}\;,\\\label{eq:extinction2}
	\kappa_s(\omega)&=&\frac{c}{\omega l}\left\{\ln\left[\frac{4n_s(\omega)n_0}{(n_s(\omega)+n_0)^2}\right]-\ln|H(\omega)|\right\}\;.
\end{eqnarray}
\end{subequations}
Thus, the variances of the refractive index and the extinction coefficient due to the magnitude and phase variances are
\begin{subequations}\label{eq:variance_amplitude}
\begin{eqnarray}\label{eq:variance_amplitude_n}
	s_{n}^2(\omega)&=&\left(\frac{c}{\omega l}\right)^2s_{\arg( H)}^2(\omega)\;,\\
	\label{eq:variance_amplitude_k}
s_{\kappa}^2(\omega)&=&\left(\frac{c}{\omega l}\right)^2s_{\ln|H|}^2(\omega)+\left[\frac{c}{\omega l}\left(\frac{n_s(\omega)-n_0}{n_s(\omega)+n_0}\right)\right]^2\frac{s_{n}^2(\omega)}{n_s^2(\omega)}\;.
\end{eqnarray}
\end{subequations}
By substituting Eq.~\ref{eq:variance_trans_simplified} into \ref{eq:variance_amplitude}, the amplitude-related standard deviation in the optical constants can be expressed as
\begin{subequations}\label{eq:ax_stddev}
\begin{eqnarray}\label{eq:ax_stddev1}
s_n(\omega) &= & \sqrt{2} \frac{c}{\omega l} s_{\arg(E_{\rm ref})}(\omega)\;,\\\label{eq:ax_stddev2}
s_\kappa(\omega) &=& \sqrt{2} \frac{c}{\omega l} \left[ s^2_{\ln|E_{\rm ref}|} (\omega) + \left[\frac{c}{\omega l}\left(\frac{n_s(\omega)-n_0}{n_s(\omega)+n_0}\right)\right]^2\frac{s^2_{\arg(E_{\rm ref})}(\omega)}{n^2_s(\omega)}\right]^{\sfrac{1}{2}}\;.
\end{eqnarray}
\end{subequations}
Eq~\ref{eq:ax_stddev} concludes the derivation of the standard deviation in the optical constants.

\acknowledgments     

The author acknowledges useful discussion with Dr John O'Hara and Dr Ibraheem Al-Naib. This research was supported by the Australian Research Council Discovery Projects funding scheme under Project DP1095151.

\bibliography{2012_MK2}  
\bibliographystyle{spiebib}   

\end{document}